\documentclass{article}
\usepackage{spconf,amsmath,graphicx}
\usepackage{multirow}
\usepackage{hyperref}
\usepackage{enumitem}
\usepackage[marginal]{footmisc}
\usepackage{amssymb}
\usepackage{stfloats}
\usepackage{booktabs}
\usepackage{makecell}
\usepackage{threeparttable}
\usepackage{cite}
\ninept 

\title{The RoyalFlush System of Speech Recognition for M2MeT Challenge}
\name{Shuaishuai Ye, Peiyao Wang, Shunfei Chen, Xinhui Hu, and Xinkang Xu}
\address{Hithink RoyalFlush AI Research Institute, Zhejiang, China}
\email{\{yeshuaishuai, wangpeiyao, chenshunfei, huxinhui, xuxinkang\}@myhexin.com}
\begin{document}
%
\maketitle
\begin{abstract}

This paper describes our RoyalFlush system for the track of multi-speaker automatic speech recognition (ASR) in the M2MeT challenge. 
We adopted the serialized output training (SOT) based multi-speakers ASR system with large-scale simulation data. 
Firstly, we investigated a set of front-end methods, including multi-channel weighted predicted error (WPE), beamforming, speech separation, speech enhancement, etc., to process training, evaluation, and test sets. However, according to their experimental results, we only selected the WPE and beamforming approach as our front-end methods. 
Secondly, we made great efforts in the data augmentation for multi-speaker ASR, including adding noise and reverberation, overlapped speech simulation, multi-channel speech simulation, speed perturbation, front-end processing, etc., which brought us a significant performance improvement. 
Finally, to make full use of the performance complementary of different model architecture, we trained the standard conformer based joint CTC/Attention (Conformer) and U2++ ASR model with a bidirectional attention decoder, a modification of Conformer, to fuse their results. 
Compared with the official baseline system, our system got a 12.22\% absolute Character Error Rate (CER) reduction on the evaluation set and 12.11\% on the test set.
\end{abstract}


\section{Introduction}
\label{sec:intro}
Although ASR has achieved superior performance in many tasks \cite{2019State,2019SpecAugment}, meeting transcription is still considered as one of the most challenging tasks \cite{2021Large}. 
Due to the distance between the speaker and microphone \cite{2006Analysis}, Meeting speech often contains noise and reverberation. 
In the meantime, multiple speakers talking in a free speaking style cause overlapped speech, and the sentences are usually less grammatical in verbal communication \cite{2021Large}. 
What is more, the number of speakers cannot be determined. 
These problems drastically decrease the accuracy and robustness of the ASR system, especially in real-world applications.

Far-field speech with complicated noise and reverberation in large meeting rooms is complex in the acoustic domain. 
Many researchers have made great efforts in improving the accuracy and robustness of far-field ASR from various aspects.
Front-end processing techniques play an essential role in multi-channel ASR, such as WPE dereverberation \cite{2012Generalization} and various beamforming methods (MVDR\cite{2009On}, neural network beamformer \cite{drude2018integrating}).
In addition, data augmentation techniques also play a crucial role in addressing data sparsity issues and the robustness of the model for far-field ASR \cite{2019SpecAugment,2015Audio,2021Data}.
Some researches pay more attention to data augmentation based model training \cite{2020Bandpass,9053169,2020The}.
In building ASR models, the front-end speech enhancement and the back-end ASR has been combined in some studies \cite{2018Multi,omalley2021conformerbased}.

Overlapped speech is another common phenomenon in multi-talker meeting, and it has a great impact on the accuracy of ASR. 
Many researches have made tremendous efforts towards overlapped ASR. 
A standard solution is to construct a cascaded system with speech separation and ASR \cite{kanda2021comparative,2019Acoustic}. 
However, due to the inconsistency between the objectives of multiple modules and the training data, the performance of the cascaded system tends to decline in the real world. 
Permutation invariant training(PIT) has been introduced into the multi-speaker ASR since it can consider all possible permutations of speakers \cite{2017Recognizing,2021Streaming}. 
Although PIT-based ASR system works well for overlapped speech recognition, the output layer of the model is limited to the number of speakers. 
It cannot handle the dependency among utterances of multiple speakers\cite{2021Unified,2020Serialized}. 
In recent years, the SOT method was proposed for solving the overlapped ASR by introducing a special symbol to represent the speaker change \cite{2021Unified,2020Serialized,2020Joint}. 
It can solve the disadvantage of PIT and naturally model the dependency among the outputs for multiple speakers.

In this work, we adopted SOT based multi-speakers ASR system.
In order to make full use of the performance complementary of different model architecture, we built two acoustic models of ASR system with the joint CTC/Attention End-to-End (E2E) framework: U2++ \cite{2021WENET} and Conformer \cite{2020Conformer}. we employed various data augmentation techniques for ASR. 
(1) We first use the single-channel data to simulate large multi-channels data. 
(2) The single-speaker data from near and far-field data is used for simulating large multi-talker speech data. 
(3) Based on two simulated data and real training data, we exploited strategies of front-end data processing methods to process multi-channel data, such as beamforming, WPE, and simulated equalization (EQ). 
(4) After that, we also used conventional data augmentation methods for all data, such as adding noise and reverberation, speed perturbation, pitch perturbation, and spectrum augmentation. 
Using these augmentations, We obtained about 18,000-hour effective training data. 
In addition, an LSTM based language model was also trained for decoding. 
Finally, we employed the ROVER tool to fuse the multiple results. Comparing with the official baseline system, the fusion result achieved a 12.22\% absolute CER reduction on the evaluation set, from 29.7\% to 17.48\% and 12.11\% on test set, from 30.9\% to 18.79\%.

The rest of the paper is organized as follows. 
In Section 2, we will introduce the M2MeT challenge and the details of data preparation for the challenge. 
Section 3 presents our speech recognition system. 
The experiments and results are shown in Section 4. 
Finally, Section 5 concludes the paper and discusses the future work.

\section{M2Met challenge and Data Preparation}
\label{datapreparation}
In this section, we first introduce the Multi-channel Multi-party Meeting Transcription (M2Met) challenge, which we participated in, 
and its dataset, namely AliMeeting and Aishell-4.
Then, we will describe our data augmentation methods, including speed perturbation, adding noise and reverberation, pitch shifting, multi-channel far-field speech simulation, overlapped speech simulation, spectrum augmentation and simulated equalization. 
%
\subsection{M2Met challenge}
\label{M2Met}
In order to inspire research on advanced meeting rich transcription, Alibaba Group launched the M2MeT challenge, as an ICASSP2022 Signal Processing Grand Challenge \cite{2021M2MeT}.
The challenge consists of two tracks, namely speaker diarization and multi-speaker ASR. 
For the track of multi-speaker ASR, the organizer also set up two subtrack: the first sub-track limits the data usage while the second sub-track allows the participants to use extra constrained data.
This work describes our submitted system to the challenge, which is only for the first subtrack of the multi-speaker ASR.
\subsection{Data preparation}

In the M2MeT challenge, AliMeeting \cite{2021M2MeT} dataset was presented along with Aishell-4 \cite{2021AISHELL} as the training data.
The AliMeeting dataset is a sizeable Mandarin meeting corpus containing 118.75 hours of real meeting data recorded by 8-channels directional microphone array and headset microphone. The detailed recorded configuration refers to \cite{2021M2MeT}. The Aishell-4 dataset, an 120 hours open-source dataset for speech enhancement, separation, recognition, and speaker diarization in conference scenarios, covers a variety of aspects in real-world meetings, including diverse recording conditions, various number of meeting participants, various overlap ratios noise types, and meets the modern requirement of accurate labeling. In the training phase, the AliMeeting is divided into 104.75 hours for training, 4 hours for evaluation, and 10 hours as a test set, and the Aishell-4 is divided into 107.5 hours for training, and 12.72 hours for evaluation.

\subsection{Data Augmentation}
Data augmentation plays an important role in model training, especially in the low resource scenario, and it can significantly increase data diversity and quantity. 
Our data augmentation methods are as follows:

 \begin{figure*}[t]
    \centering
    \includegraphics[width=\linewidth]{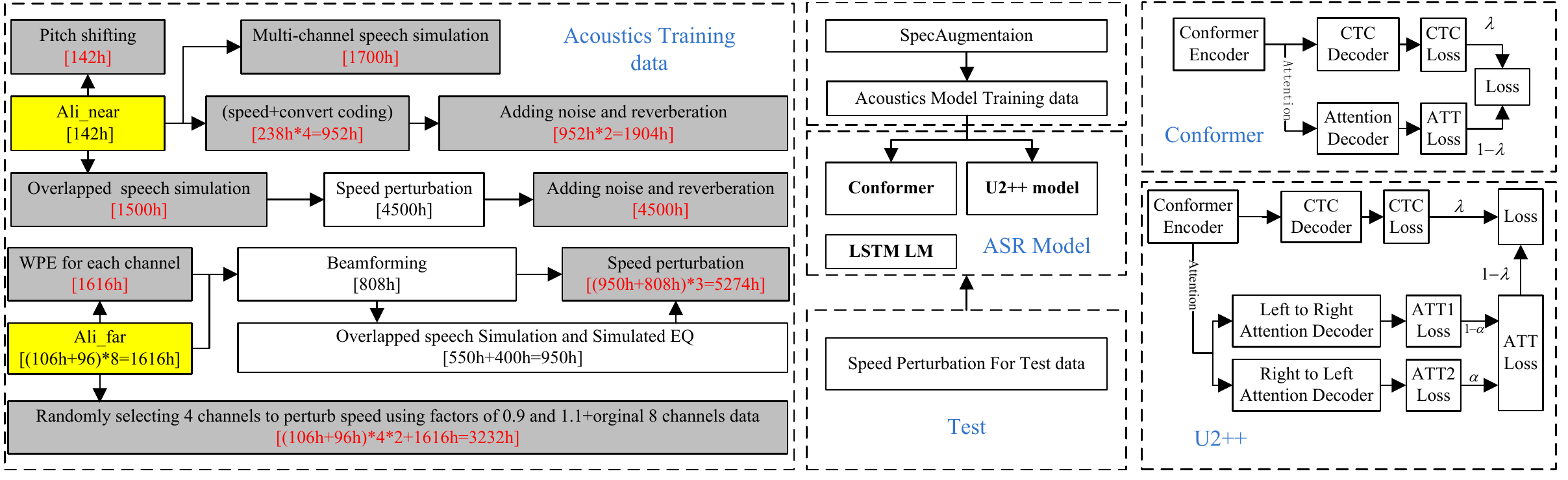}
    \caption{The left part is the acoustics training data, where we listed our all data augmentation methods, the yellow blocks are original data set and the grey blocks were used to train our models. In the training stage, the utterances greater than 30 seconds were excluded, so the final duration of training set is about 18,000 hours. The middle part is our system architecture, including training data, ASR model and test data processing. The right part is two ASR model architectures.}
    \label{fig:m2met_all}
\vspace{-1.8em}
\end{figure*}

\begin{itemize}[leftmargin=*]
    \item \textbf{Speech perturbation}: Speech perturbation\cite{2015Audio} 
    is an audio-level speech augmentation method. In this work, It produces 3 variants of the original signal with speed factors of 0.9, 1.0, and 1.1.

    \item \textbf{Adding noise and reverberation}: Adding noise and reverberation to corrupt clean speech can simulate the noisy and reverberant conditions for some  rooms. In our system, the noise sources are composed of MUSAN corpus \cite{2015MUSAN}, excluding bable noise and AliMeeting noise source extracted from the AliMeeting training set according to timestamp. The reverberation sources include two parts, RIRS\_NOISES corpus \cite{2017RIRS} and AliMeeting single-channel reverberation source simulated based on image method \cite{1998Image} according to the room sizes of AliMeeting using a public open-source  code\footnote{https://github.com/ehabets/RIR-Generator}.

    \item \textbf{Pitch shifting}: One copy was created for each original data by modifying the pitch.
    We used the \textit{Sox} audio manipulation tool \cite{sox} to realize pitch shifting. 
    
    \item \textbf{Multi-channel far-field speech simulation}: 
    For speech data captured by a headset microphone, we conducted a multi-channel data simulation to make full use of it under the far-field environment. 
    Concretely speaking, we generated 17,400 multi-channel room impulse responses (RIRs) based on the configuration of the microphone array with the image method\cite{1998Image}. 
    2400 of these RIRs were generated according to the AliMeeting room information, and the others were generated according to random room size, which ranges from $3*3*2.5m^3$ to $10*10*3.5m^3$. 
    Then we took the headset data as the sound source, converted it into 8-channel far-field data with the simulated RIRs, and added noise from different directions to simulate the array data. 
                                            
    \item \textbf{Overlapped speech simulation}:
    According to \cite{2021M2MeT},  
    the training and evaluation data's average speech overlap ratio is 42.27\% and 34.76\%, respectively.
    In order to improve ASR model performance for recognizing the  overlapped speech, we employed the non-overlapped speech in the data sets recorded by an 8-channel directional microphone array and headset microphone to simulate a large amount of overlapped speech data with proportions of 50\% for 2 speakers, 30\% for 3 speakers, 20\% for 4 speakers, respectively.
    
    \item \textbf{Spectrum augmentation (SpecAugment)}: 
    For improving the model performance and the robustness of our ASR model, we applied the SpecAugment method\cite{2019SpecAugment}  to all input features of a neural network at a mini-batch level. The SpecAugment includes warping the features, masking blocks of frequency channels, and masking blocks of time steps, and we only utilized the last two tricks in our system. We set the parameters of SpecAugment to default values in WeNet \cite{2021WENET} and EspNet \cite{2018ESPNET} tools.

    \item \textbf{Simulated equalization(EQ)}: 
    Inspired by \cite{ren2021causal}, we used various filters to process data, including weighted low-pass/high-pass filter, de-emphasis filter, simulated frequency response curve filter and so on.
    
\end{itemize}

\section{System Description}

In this section, we will elaborate on our ASR system  submitted to the M2MeT challenge. 
The system diagram is shown in the Figure \ref{fig:m2met_all}. 
Our ASR system mainly consists of front-end processing, data preparation, acoustic model, language model, and model fusion.
For the reason that data augmentation has been introduced in section\ref{datapreparation}, here we will describe the remaining parts.

\subsection{Front-end processing}
The WPE\cite{2012Generalization} has been shown to be an effective method for speech dereverberation in far-field scenarios, thus improving the recognition performance. It is a kind of long-term linear prediction method to estimate the reverberation tail in the speech signal and subtract it from the observed signal. The original assumption of the WPE method is that there is no noise and overlapped speakers. However, in our experiments, it is found that the WPE can improve the performance of the speech recognition system even in an environment with additive noise and overlapped speech. The same conclusion was also obtained in \cite{drude2018integrating}.

After the WPE was used for dereverberation, the weighted delay-and-sum beamforming and scheme reference from the beamformIt\cite{2007Acoustic} were further utilized for processing.
We select a reference microphone by calculating the crosscorrelations of two microphone pairs. Then it uses Generalized Cross Correlation with Phase Transform (GCC-PHAT) method to calculate the time delays of arrival(TDOA) between each microphone and the reference microphone. Furthermore, the best TDOA is estimated by a 2-step Viterbi search method. 
Finally, the signals are aligned and summed by each channel's estimated TDOA and weights.

\subsection{Acoustic model}
Different model architectures have different modeling capabilities for ASR, so they can complement each other to improve system performance. 
For this reason, we adopted two model architectures in our system, namely the standard Conformer based joint CTC/Attention ASR model (Conformer) \cite{2020Conformer} and the Conformer based U2++ ASR model\cite{wu2021U2++}, as shown in the right part of Figure \ref{fig:m2met_all}.

The Conformer consists of: a shared encoder, an attention decoder, and a CTC module seeing the right-top part of Figure\ref{fig:m2met_all}. 
The U2++ model has been made two modifications based on the Conformer mentioned above;
one is to modify the attention decoder into a bidirectional decoder, seeing the right-bottom part of the Figure\ref{fig:m2met_all} and the other is to propose attention rescoring, a new decoding mode. 
The more descriptions for Conformer and U2++ refer to \cite{2020Conformer} and \cite{wu2021U2++}.

\subsection{Language model}
For the Conformer model, a recurrent neural network language model (LM) was trained by the Espnet\cite{2018ESPNET}. %
The training transcripts consist of two parts: One part is the original training data, and the other one is an augmentation text by concatenating several consecutive sentences\cite{2020Context}. 
The model architecture consist of 2 LSTM  layers with 1024 dimensions and the model unit is character.
\subsection{Model fusion}
To fuse the above mentioned ASR systems, the fusion strategy Rover \cite{2002A} is used.
It is a system developed at NIST to produce a composite ASR system output when the outputs of multiple ASR systems are available. The Rover system implements a voting or rescoring process to reconcile differences in ASR system outputs. 

We prepared the candidate results from U2++ and conformers without/with LM and performed speed perturbation to the test set with a speed factor of 0.9 and 1.1, respectively.

\section{Experiments and Results}
\label{sec:typestyle}
\subsection{Experimental Setup }

The results of M2MeT are summarized in \cite{Yu2022Summary}, where our system is denoted as 'R62'. Our experiments and results are summarized as follows.

All our experiments were conducted on the EspNet for the Conformer and the WeNet for U2++ model.
The acoustic feature of 83 dimensions, 80-dimensional Fbank plus the 3-dimensional pitch feature was extracted from every frame with a frame length of 25ms and a frame shift of 10ms.

The standard conformer based on joint CTC/Attention  model is composed of a 12-layer encoder with 2048-dim feed-forward and depth-wise convolution with kernel size 3, a 6-layer decoder with 2048 units and 8 heads attention with 512 dimensions. 
The U2++ model was configured with the same settings as the aforementioned standard model, expect for an 18-layer encoder and modifying a 6-layer decoder into a 3-layer left-to-right decoder and a 3-layer right-to-left decoder.
Both models employed the characters in training data as the modeling unit. We set both the CTC weight and the reverse weight to 0.3 in the training phase and set 0.3 and 0.6 respectively in the decoding phase. 
The beam size was set to 20. We trained 30 epochs for the U2++ model and 11 epochs for the  Conformer model with the same learning rate of 0.001.  During the training stage of the Conformer model, the first 10 epochs is trained using 10000 hours and the last epoch is trained using 14000 hours of the acoustics training data.
We averaged the best 10 and the best 8 models during the decoding stage based on the loss value as the final model for the U2++ and Conformer models respectively.
\begin{table*}[th]
  \caption{The results of different models on the  AliMeeting far-field evaluation and test set. 'wb' refers to 'wpe-bf', and 'Eval' refers to evaluation set. 'U2++\_*channel' represents the number of channels. 'U2++\_big\_data' represents the final model trained by all data.}
  \setlength{\tabcolsep}{0.8mm}{
  \label{tab:results}
  \centering
  \begin{threeparttable}
  \begin{tabular}{cccccccc|ccc}
    \toprule
      \multirow{2}{*}{Group}   & \multirow{2}{*}{System}       &   \multicolumn{6}{c|}{Eval (CER\%)}                                                         & \multicolumn{3}{c}{Test (CER\%)} \\
      \cmidrule{3-11}
                {}             & {}                            &  {Eval-1c}  & {Eval-wpe} & {Eval-bf} & {Eval-wb}          & {Eval-wb-sp0.9}    & {Eval-wb-sp1.1}    & {Test-wb}           & {Test-wb-sp0.9}      & {Test-wb-sp1.1}  \\
                
    \midrule
             {Group1}          & {baseline\cite{2021M2MeT}}    &  {30.8}    &  {-}      & {29.7}   & {-}               & {-}               &          {-}      & \multicolumn{3}{c}{30.9}           \\
    
    \midrule
       \multirow{5}{*}{Group2} & {Conformer\_Ali-near}         & {47.44}    & {-}       & {-}      &  {-}              & {-}               & {-}               &       {-}           &        {-}           &   {-}     \\
                 {}            & {+SOT}                        & {46.06}    & {-}       & {-}      &  {-}              & {-}               & {-}               &       {-}           &        {-}           &   {-}     \\
                 {}            & {Conformer\_5k-hours}         & {24.02}    & {24.99}   & {23.32}  &  {22.77}          & {-}               & {-}               &   {23.15}           &        {-}           &   {-}   \\
                 {}            & {Conformer\_10k-hours}         & {21.49}    & {21.71}   & {20.27}  &  {\textbf{19.06}} & {\textbf{19.74}}  & {\textbf{19.31}}  &   {\textbf{20.14}}  &    {\textbf{20.65}}  &   {\textbf{20.68}}  \\
                 {}            & {+LSTM LM}                    & {21.64}    & {21.84}   & {20.52}  &  {\textbf{19.14}} & {19.90}           & {19.59}           &   {\textbf{20.29}}  &       {-}            &   {-}   \\
    \midrule
      \multirow{4}{*}{Group3}  & {U2++\_1channel}              & {28.49}    & {28.95}   & {26.83}  &  {26.13}          & {-}               & {-}               &  {26.78}            &       {-}            &   {-}     \\

                 {}            & {U2++\_4channels}             & {27.93}    & {28.47}   & {26.56}  &  {26.04}          & {-}               & {-}               &  {26.64}            &       {-}           &   {-}  \\

                 {}            & {U2++\_8channels}             & {27.45}    & {27.86}   & {26.15}  &  {25.50}          & {-}               & {-}               &  {26.20}            &       {-}           &   {-}    \\
 
                 {}            & {U2++\_big\_data}             & {21.10}    & {21.08}   & {19.92}  &  {\textbf{18.68}} & {\textbf{18.82}}  & {\textbf{19.12}}  &  {\textbf{19.99}}   &  {\textbf{20.10}}   &   {\textbf{20.46}}  \\
    \midrule 
                 {Group4}      & {Fusion*}                     &  \multicolumn{6}{c|}{\textbf{17.48}}                      &  \multicolumn{3}{c}{\textbf{18.79}}               \\

    \bottomrule
  \end{tabular}
\begin{tablenotes}
\footnotesize
\item[*] The numbers with bold font are fused as the final result using the \textit{ROVER} tool. 
\end{tablenotes}
\end{threeparttable}}
\vspace{-1.5em}
\end{table*}

\subsection{Results and analysis}
Due to the different hyper-parameters of the two models trained in our system, their results are not comparable. In this section, we mainly conducted different experiments on two models, respectively. Our experiments were conducted on the following several aspects to investigate the influence on the ASR model:
(1) SOT scheme;
(2) the amount of training data;
(3) the number of the channel used to train model;
(4) front-end methods, namely the WPE and beamforming approach;
(5) model fusion.

\subsubsection{SOT scheme}
In the experiments of the conformer model, we verified the effectiveness of the SOT method on the same duration of the simulated overlapped near-field speech data Ali-near. The results on the evaluation set are shown in first two lines of Group2 in Table \ref{tab:results}. Compared with a model trained using the Ali-near, we can find that the SOT method can significantly improve the performance. Therefore, the following experiments were conducted by adopting the SOT method. More detailed experimental verification for SOT refers to \cite{2020Serialized}.
\subsubsection{The number of channel for training model}
In order to verify whether the number of channel used to train the model will affect the model's performance, we carried out three experiments on U2++ ASR model, and the results are shown in the first three lines of Group3 in Table \ref{tab:results}. With the increasing number of channels, we can see that the performance has also been improved to varying degrees.  We inferred that the gain is mainly caused by channel differences which can improve the robustness of the ASR model. So, we used 8-channel far-field data to simulate overlapped speech and trained the ASR models in the following experiments.

\subsubsection{The amount of training data}
With the increase of training data, the performance of the ASR model generally becomes improving. 
We conducted 2 experiments with a 5000-hour and a 10000-hour of training data for the Conformer model as shown in the Group2 of Table \ref{tab:results} and 4 experiments in the Group3 for U2++ model. Compared with the baseline model \cite{2021M2MeT}, we find that the conformer model and the U2++ model achieved 9.31\% and 9.7\% absolute CER reduction on the Ali-1c without front-end data, and 10.61\% and 10.91\% absolute CER reduction  on the test set, respectively. The significant performance improvement was mainly attributed to the data augmentation. Based on these facts, we see that a more considerable amount of training data can obtain better performance. Therefore, we can conclude that large-scale training data with a set of data augmentation methods can effectively improve the performance and robustness of the model.

\subsubsection{Front-end}
Since the WPE can only reduce the reverberation tail, its benefit to the improvement of the ASR performance is not so large.
The beamforming can mask part of noise and reverberation relying on spatial information, and it is capable of improving the signal-to-noise ratio (SNR) in the speaker direction. 
So, by combining the WPE and beamforming methods, both their advantages can be utilized. 
As shown in Table \ref{tab:results}, the beamforming can significantly decrease the CER of the evaluation and test set. 
Furthermore, the performance can be prominently improved when WPE and beamforming are used successively. 
\subsubsection{Model fusion}
In order to further reduce the CER on the evaluation set, we fused 7 subsystems with ROVER tool, as the bold font in Table \ref{tab:results}. Compared with the best single system, the fusion scheme brought a CER absolute reduction of 1.2\% from 18.68\% to 17.48\% on the evaluation set and 1.2\% from 19.99\% to 18.79\% on the test set. The possible reasons for the improvement were summarized as follows: 
  (1) With the perturbing processing, the incorrectly recognized characters caused by those speeches that are too fast or too slow can be recognized correctly.
  (2) As a supplementary result to the original utterance, the LSTM based language model can correct some syntax error.
\section{Conclusion}
This paper described our multi-speaker ASR system for the M2MeT challenge. 
Two E2E ASR frameworks, Conformer and U2++, were adopted with SOT method. 
Various front-end processing methods, data augmentations, model fusion, and LSTM LM, were investigated. 
Experimental results showed that the WPE and beamforming based on multi-channel can effectively decrease the CER.
We found that using a variety of data augmentations to expand the training set can  significantly improve performance and robustness. 
We also found that the fusion of different models and speed perturbation can achieve 1.2\% absolute CER reduction on the evaluation set compared with the best result of the U2++ ASR model.
Comparing with the official baseline system, our system got a 12.22\% absolute CER reduction on the evaluation set, from 29.7\% to 17.48\% and 12.11\% absolute CER reduction on the test set, from 30.9\% to 18.79\%.


\vfill\pagebreak


\bibliographystyle{IEEEbib}
\bibliography{strings,refs}

\end{document}